\documentclass[11pt, oneside]{article}   	% use "amsart" instead of "article" for AMSLaTeX format
\usepackage{geometry}                		% See geometry.pdf to learn the layout options. There are lots.
\usepackage{array}
\usepackage{graphicx}	
\usepackage{subcaption}
\usepackage{amssymb}
\usepackage{hyperref}
\hypersetup{
    colorlinks=true,
    linkcolor=black,
    filecolor=black,      
    urlcolor=black,
    citecolor=black,
    pdftitle={},
    pdfpagemode=FullScreen,
    }
    
\usepackage{color}
\usepackage{tabularray}
\usepackage[font=small,labelfont=bf]{caption}
\usepackage{subcaption}
\usepackage{booktabs}
\usepackage{wrapfig}
\usepackage{siunitx}
\usepackage{arydshln}

\newtheorem{criteria}{Redistricting Criteria}[section] %the resolution could also be [subsection]

\renewcommand{\sp}{\mathrm{sp}}

\title{A retrospective analysis of Montana's 2020 congressional redistricting map
}
\author{Kelly McKinnie and Erin Szalda-Petree}
\begin{document}
\maketitle

\begin{abstract}
    The 2020 decennial census data resulted in an increase from one to two congressional representatives in the state of Montana. The state underwent its redistricting process in 2021 in time for the November 2022 congressional elections, carving the state into two districts. This paper analyzes the redistricting process and compares the adopted congressional map to the space of all other possible maps. In particular, we look at the population deviation, compactness and political outcomes of these maps. We also consider how well two popular sampling techniques, that sample from the space of possible maps, approximate the true distributions of these measures.
\end{abstract}

%\section{}
%\subsection{}
\section{Introduction to Redistricting in Montana}
\label{sec:intro}
\begin{figure}[h!]
\begin{subfigure}[t]{.48\linewidth}
\includegraphics[width=\linewidth]{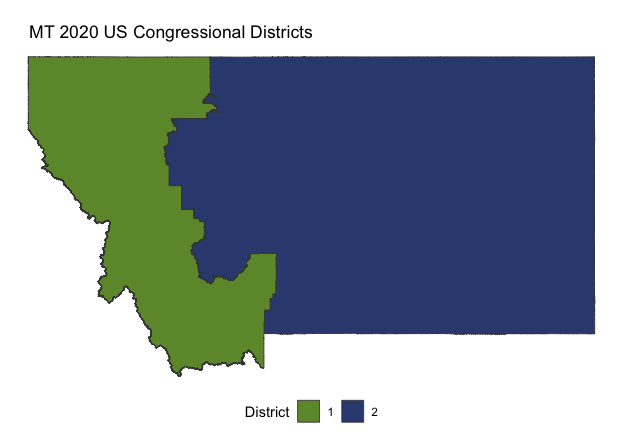}
% \caption{}
% \label{fig:FinalMap}
\end{subfigure}
\hfill
\begin{subfigure}[t]{.48\linewidth}
\includegraphics[width=\linewidth]{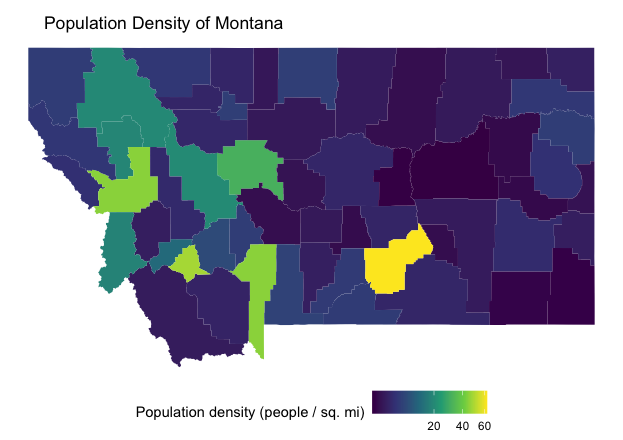}
% \caption{}
% \label{fig:PopDensity}
\end{subfigure}
\caption{Montana's two adopted congressional districts and population density by county.}
\label{fig:MT}
\end{figure}

According to the \href{https://www.census.gov/data/tables/time-series/dec/popchange-data-text.html}{US decennial census}, between 2010 and 2020 Montana's population grew from 989,415 to 1,084,225, an increase of 94,810 people. This difference, along with overall US demographics, was enough to push the number of US congressional districts in Montana from one to two. Montana's five member \href{https://leg.mt.gov/districting/}{Districting and Apportionment Commission} (henceforward referred to as DAC) was constitutionally tasked with drawing the new districts. The DAC consists of one commissioner selected by the majority and minority leaders of each house of the Legislature. The 5th commissioner is elected by the first four within 20 days or appointed by the Montana Supreme Court if consensus cannot be reached. On November 12, 2021 the DAC submitted the \href{https://leg.mt.gov/content/Districting/2020/Maps/Congressional/Adopted/MT-Congressional-Districts-adopted-Nov12.pdf}{adopted plan} to the Secretary of State's office. A map of the two districts is given in Figure \ref{fig:MT} along with a population density by county map.

Before choosing the final map, the DAC asked for public input in fall of 2021, eventually putting 12 maps forward for further consideration. The redistricting process invites several questions. How many ways are there to divide the state of Montana into two congressional districts? How many of these maps can ``reasonably'' be considered? How does the adopted map compare to the maps in the set of all possible maps with respect to differences in population, geographic compactness, political leanings? In this paper we will study these questions using both enumeration techniques developed in \cite{enumpart} and sampling techniques developed in \cite{smc}, found in the R package \href{https://cran.r-project.org/web/packages/redist/index.html}{redist}, and using the python package \href{https://mggg.github.io/GerryChain/}{Gerrychain} with the MCMC techniques developed in \cite{Recomb}.

Though technically the DAC could create districts afresh using the 88,417 Montana census blocks, they chose to create districts that nearly followed county boundaries. The Congressional map submitted by the DAC splits only one of the 56 Montana counties; Pondera (Pahn-dur-ay). At the last DAC meeting Pondera was split down the middle to achieve maximum population equality, reversing a previously agreed upon amendment where the split was along the Blackfeet Indian Reservation  \cite{MTPR1} and increasing criticism that the Native American population is diluted by the map \cite{MTPR2}. In our analysis we will use the ``57 county map'' which consists of the 55 standard counties plus E Pondera and W Pondera, so that the adopted map lies among the set of possible maps. 

\section{From one to two districts}
\label{sec:one_to_two}
How many ways are there to redistrict the state assuming that we split only along county lines? There are 56 counties in Montana. If there are no constraints on how to form the districts, then there are $2^{55} = 36,028,797,018,963,968$ (about 36 quadrillion) ways to partition the 56 counties into two districts, and twice as many if we allow the Pondera split. However, as is natural and common, the DAC adopted a set of criteria based on the Montana Constitution as well as state and federal laws to restrict the drawing of new districts. These can be found on the \href{https://mtredistricting.gov}{DAC website}. Mandatory criteria are set out by law, while the discretionary criteria are traditional redistricting principles selected by the DAC to provide further guidance on where to draw lines. 

\begin{figure}[h!]
  \centering
  \begin{subfigure}[t]{0.48\linewidth}
  \includegraphics[width=\linewidth]{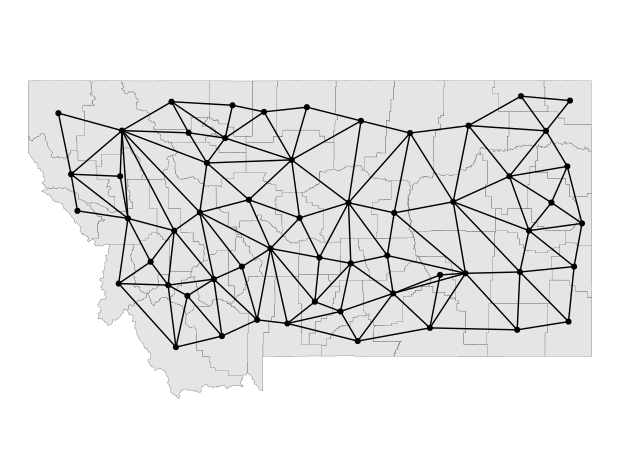}
  \caption{MT\_141. The full MT graph has 141 edges.}
  \label{fig:fullmt}
  \end{subfigure}
  \hfill
  \begin{subfigure}[t]{.48\linewidth}
  \includegraphics[width = \linewidth]{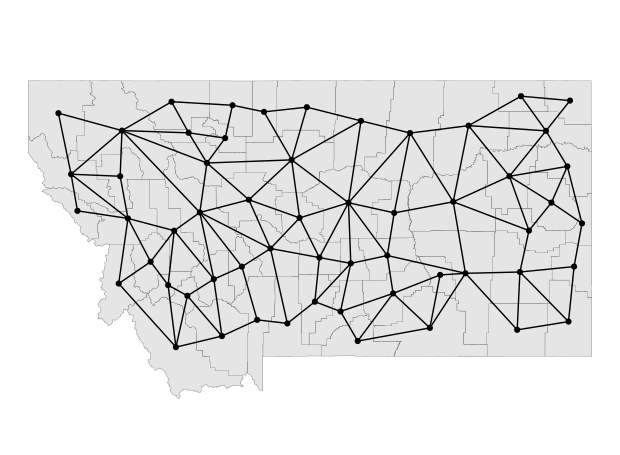}
  \caption{MT\_122. Reduced MT Map with edges eliminated if shared perimeter is $<38$km and $<$ 10\% of each county. 122 edges.}
  \label{fig:122mt}
  \end{subfigure}
  \caption{MT counties shown with graphs.}
 
\end{figure}

The first criteria that we consider here is about contiguity.
\begin{criteria}
Each district shall be contiguous, meaning that a district must be in one piece. (Article 5, Section 14 of the Montana Constitution). Areas that meet only at points of adjoining corners shall not be considered contiguous. Areas separated by natural geographical or artificial barriers that prevent transportation by vehicle on a maintained road shall be avoided when not in conflict with the commission’s adopted criteria and goals.
\label{c1}
\end{criteria}
To mathematize this idea, the counties in Montana are turned into a graph; each node in the graph represents a county and two nodes are connected by an edge if their respective counties share a border. This is known as the `dual graph' to the map. Public Law 94-171 requires the Census Bureau provide states with geographic and census data to redraw district lines. The Census Bureau provides geographic shape files from which the graph can be computed. From the census generated shapefiles the package {redist} \cite{redist} in R  plots the MT counties with the adjacency graph as an overlay as in  Figure \ref{fig:fullmt}. This graph has 57 nodes and 141 edges. We will refer to this graph as MT\_141 when needed.% (but do we ever refer to it?).

A close look at the map in Figure \ref{fig:fullmt} shows that there are counties which are connected by a very short boundary. Though technically meeting the standard of contiguity, it is unlikely that the DAC would have placed two counties connected by a very short border in the same district only because of that border. If they were, the shape of the congressional district would likely fail to be compact enough for consideration (we address compactness in Section \ref{sec:compact}). For this reason we eliminate edges of the graph which correspond to small shared borders. In particular, we eliminate edges that correspond to shared borders with length $<38$ km and comprise less than 10\% of each county's border. Note that this does not prevent the counties with eliminated short borders from residing in the same congressional district. It just means that they won't be in the same district only because of that short border. Table \ref{table:borderlengths} in the Appendix contains the list of shared perimeter lengths in increasing order with length $< 50$ km.

There are 19 edges with shared perimeter length $<$ 38 km and $<10$\% of each county border. Eliminating these leaves 122 edges in the graph which we call MT\_122. The 57 county map with MT\_122 overlaid is shown in Figure \ref{fig:122mt}.

When a district map is drawn along county lines it partitions the counties into two subsets. 
To understand how the adopted map compares to all possible maps one would like to 
\begin{wraptable}{l}{.48\textwidth}%[h!]
\centering
\begin{tabular}{|p{.7in}|p{.6in}|p{1in}|}
\hline
Edges 

Removed & Edges in 

Graph & \# Redist

Solutions\\
\hline\hline
None &\centering\vfill141&\vfill11,976,688,820\\

\hline
$<10\%$ \& $<38$ km &\centering\vfill 122 & \vfill 495,691,401\\
\hline
\end{tabular}
\caption{Number of redistricting solutions}% - updated 5/31/23}
\label{table:enumpart}
\end{wraptable}
know all possible ways to partition the county map into two subsets which abide by the contiguous criteria (\ref{c1}). We call each allowable partition a {\it redistricting solution}. In \cite{enumpart}, Fifield, et al. give an algorithm for doing exactly this; enumerating the partitions which result in exactly two contiguous components (more generally the algorithm will enumerate partitions into $n$ contiguous components for any $n>1$). Using the associated function {\it redist.enumpart}, in the R package {redist} \cite{redist},  there are about 11.2 billion ways %7,012,136,966 ways
to redistrict Montana into two districts along the 57 county lines (including the Pondera split) using MT\_141. Using MT\_122 there are 495,691,401 %4,996,925,074 
redistricting solutions. These results are summarized in Table \ref{table:enumpart}. 

Though it took only seconds to compute the number of solutions in both the full and the restricted graph, it took almost a full day to record the roughly 500 million redistricting plans from the restricted graph. Though technically feasible, it was impractical to record the full set of solutions from MT\_141. Since the edges removed were so short, we are confident that we did not eliminate redistricting solutions which would have been considered reasonable by the commission and thus they are a functionally complete set of solutions with which to compare the adopted map. Our solutions do not, however, include the possibility of a different county being split or, more generally, maps drawn not with respect to county lines.

\section{Population Criterion}
\label{sec:pop}
Another criteria that the DAC considers is the population of each district:
\begin{criteria}
    Districts must be as equal in population as is practicable (Article 1, Section 2, U.S.
Constitution).
\end{criteria}

Given a geographic region with population $p$ which is to be split into $n$ districts we call $\bar p = p/n$ the {\bf ideal population} of each district. If district $i$ has population $p_i$, the {\bf population deviation} of district $i$ is defined as $d_i = |1-p_i/\bar p| = |\bar p - p_i|/\bar p$. Population deviation measures how far a district's population is from ideal. In general one could study $\max(d_i)$ or $\max(p_i-p_j)/\bar p$ for the maximum population deviation among all districts. However, when $n=2$, as in Montana's congressional districts, $p = p_1+p_2$ gives  $1-p_1/\bar p = -(1-p_2/\bar p)$, so that $d_1 = d_2$. That is, the population deviation of the two districts are equal and we can talk about {\it the population deviation} of a redistricting plan.

In the 2020 Census, the state of Montana had 1,084,225 people. %Let $p$ be this value. 
The ideal district population is therefore $\bar p=\frac12p = 542,112.5$. A population deviation restriction of 1\% or 0.01 is a max difference of 5,421.125 people from ideal. That is, if $p_i$ is the population of district $i$, then $p_i$ must satisfy $536,691<p_i<547,533$. 

As per the criteria, the redistricting plans are filtered to contain only those those which have small population deviation. The number of redistricting solutions under different population deviation restrictions are listed in Table \ref{table:pop}.

\noindent
\begin{minipage}{\textwidth}
  \begin{minipage}[b]{0.48\textwidth}
    \centering
    \includegraphics[width = .85\linewidth]{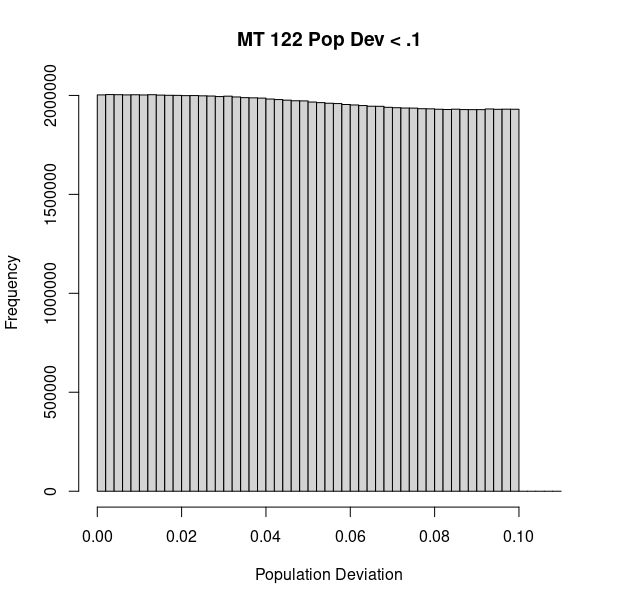}
    %\rule{6.4cm}{3.6cm}
    \captionof{figure}{Histogram of population deviations of the 98,350,198 possible plans from MT\_122 with pop\_dev $< 10$\%.}
    \label{fig:popdev}
  \end{minipage}
  \hfill
  \begin{minipage}[b]{0.48\textwidth}
    \centering
 %   \begin{table}[htp!]%{l}{.48\textwidth}

\begin{tabular}[.9\linewidth]{|p{.8in}|p{.8in}|}
\hline
Pop Dev 

Restriction &  \# Redist

Solutions\\
\hline\hline
\centering{None} &495,691,401\\
\hline
\centering{$<0.1$} & 98,350,198\\
\hline
\centering{$<0.03$} & 30,011,928\\
\hline
\centering{$<0.01$} &10,015,984\\
\hline
\end{tabular}
\vspace{.35in}
\captionof{table}{Number of redistricting solutions from MT\_122 graph with various population deviation restrictions.}
\label{table:pop}
    \end{minipage}
\end{minipage}

\medskip
To get an idea of what the distribution of population deviations looks like among the 495,691,401 possible redistricting plans, Figure \ref{fig:popdev} shows the distribution of population deviations for all plans with population deviation $<$ 0.1. Notice the nearly uniform distribution of the population deviations within 0.1.

The adopted districts have populations that differ by 1 person. This gives a population deviation of $9.2\times 10^{-7}$ (at least in theory according to the census data)! In practice, the final lines splitting Pondera were drawn to achieve this maximum population balance between the two districts. It is unreasonable to only compare plans within this extremely tight population restriction. In our analysis in the following section we will work with the less restrictive population deviation restriction of 0.03. As shown in Table \ref{table:pop}  this leaves 30,011,928 possible redistricting plans to analyze.

\section{District Compactness} 
\label{sec:compact}
In a likely effort to try to avoid gerrymandering and to improve the practical aspects of representing a district, the DAC includes the following common criteria
\begin{criteria}
Each district shall consist of compact territory. (Article 5, Section 14 of the Montana Constitution). The Commission shall consider the district’s functional compactness in terms of travel and transportation, communication, and geography.
\end{criteria}
The DAC guidelines use the word ``compact'' without a precise mathematical or geographical definition. The commission may have intended to adhere to the ``I'll know compact when I see it'' (\cite{KnowIt}) method of determining compactness. However, \href{https://leg.mt.gov/bills/mca/title_0050/chapter_0010/part_0010/section_0150/0050-0010-0010-0150.html}{Title 5, Chapter 1, Part 1} of the Montana Constitution states:
\begin{quote}
The districts must be compact, meaning that the compactness of a district is greatest when the length of the district and the width of a district are equal. A district may not have an average length greater than three times the average width unless necessary to comply with the Voting Rights Act.
\end{quote}
It is not clear what the ``average length'' of a district is. For example, is negative space included when calculating the average height? What is the average width of a rainbow or an X shape? Or a straight line at a $45^\circ$ angle? Are width and length restricted to north-south and east-west measurements, or can one consider the longest distance between two points in the district as defining one of the axes on which to measure? The average length/width definition is not standard. We could find no other states that use this measure and very few states giving such a rigid definition of compactness. Iowa is a notable exception giving a \href{https://www.legis.iowa.gov/docs/code/42.4.pdf}{very precise definition of compact}. These reasons might be why the DAC did not include this particular criteria in the 2020 session.

There are several common ways to measure compactness of a district (see e.g., \cite{MR3308268}, \cite{PolGeo}). The one we use first as a filter on the redistricting solutions is the ``Edges Removed'' (ER) measure. The ER score counts the number of edges removed in the graph to separate the two districts.  It is also referred to as the length of the edge cut. In theory, the fewer the number of edges removed, the more compact, or natural looking, the district is. 

\begin{figure}[h]
\centering
\begin{subfigure}[t]{0.3\linewidth}
\centering
\includegraphics[width= \linewidth]{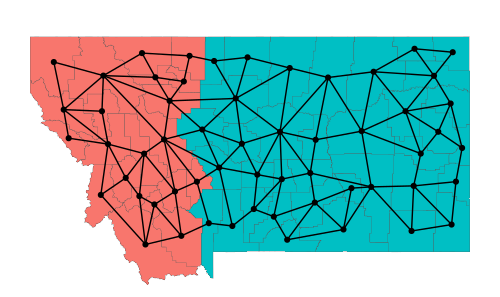}
% computed using STATS on 122 MT graph.R file on linux computer 6-1-2023
\caption{Redistricting solution with minimal ER score of 8 and population deviation of 0.027.}
\label{fig:minER}
\end{subfigure}
\hfill
\begin{subfigure}[t]{0.3\linewidth}
\centering
\includegraphics[width= \linewidth]{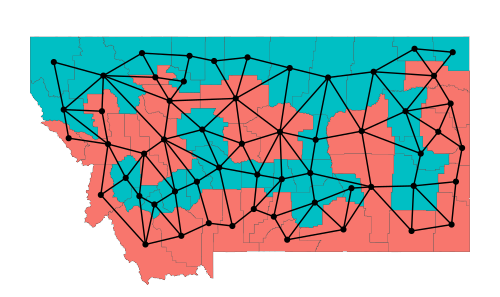}
% computed using STATS on 122 MT graph.R file on linux computer 6-1-2023
\caption{Redistricting solution with maximum ER score of 64 and population deviation of $0.0004$}
\label{fig:ERmax}
\end{subfigure}
\hfill
\begin{subfigure}[t]{0.3\linewidth}
\centering
\includegraphics[width= \linewidth]{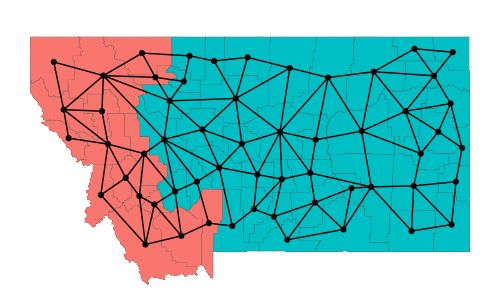}
% computed using STATS on 122 MT graph.R file on linux computer 6-1-2023
\caption{Adopted Congressional Districts (November 2021) with ER score of 11.}
\label{fig:adoptedER}
\end{subfigure}

\caption{Maps with varying ER scores within the ensemble of enumerated plans.}
\end{figure}

\begin{figure}[h]
    \centering
    \begin{subfigure}[t]{.32\textwidth}
        \centering
        \includegraphics[width = \linewidth]{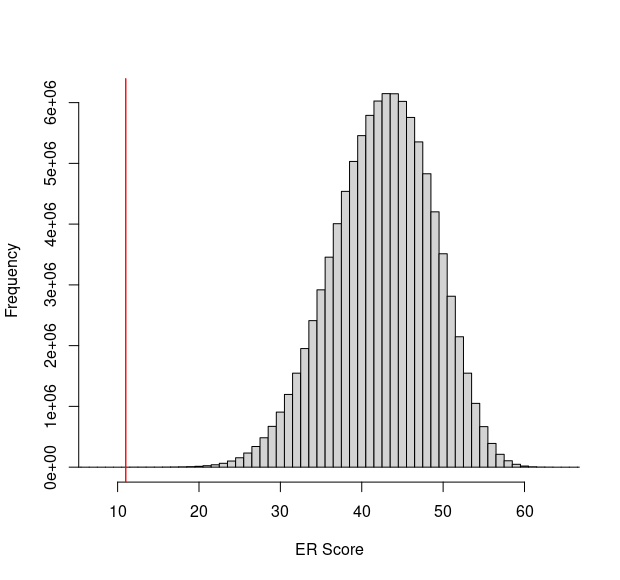}
        \caption{}
        % computed using STATS on 122 MT graph.R file on linux computer 6-1-2023
        \label{fig:ERhist}
    \end{subfigure}
    \hfill
    \begin{subfigure}[t]{.32\textwidth}
        \centering
        \includegraphics[width = \linewidth]{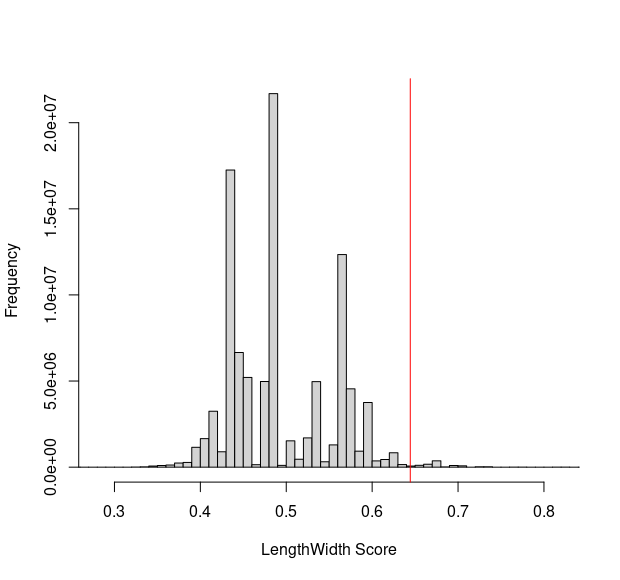}
        \caption{}
        \label{fig:lw}
    \end{subfigure}
    \hfill
    \begin{subfigure}[t]{.32\textwidth}
        \centering
        \includegraphics[width = \linewidth]{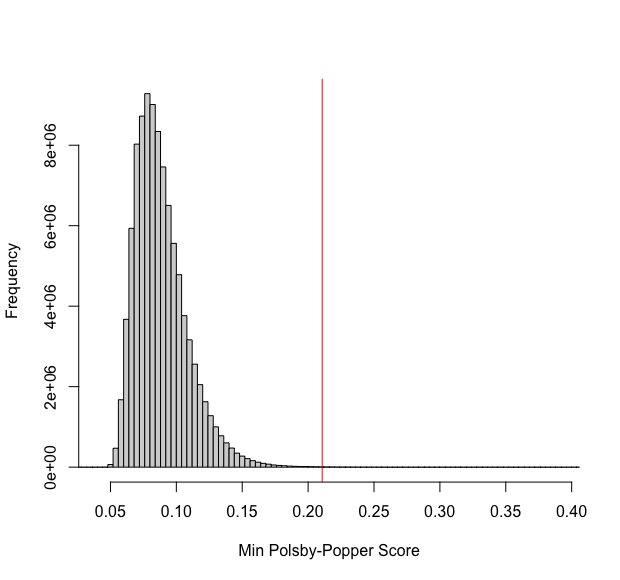}
        \caption{}
        % computed using STATS on 122 MT graph.R file on linux computer 6-1-2023
        \label{fig:pbp_hist}
    \end{subfigure}
    \caption{Histograms of ER, LW and PbP score for the 98 million plans with population deviation less than 0.1. The adopted map has ER = 11, LW = 0.64 and min PbP = 0.21.}
    \label{fig:comp}
\end{figure}

A search of the roughly 98 million redistricting solutions with population deviation $<0.1$ finds a minimum ER score of 8. There are two such plans, one is shown in Figure \ref{fig:minER}. The maximum ER score among these plans is 64 and there are 32 such plans. The one with smallest population deviation ($0.0004$) is shown in Figure \ref{fig:ERmax}. For comparison, the adopted congressional plan has an Edges Removed measure of 11 as can be seen in Figure \ref{fig:adoptedER}. The histogram of ER scores of all $<0.1$ population deviation plans can be seen in Figure \ref{fig:ERhist}. 

Our analysis in the following sections restricts to those plans with population deviation $<.03$. We will also restrict the ER score of the plans to those with ER $<22$. Table \ref{table:ER} gives the number of plans with population deviation $<.03$ and various ER restrictions.

Although the Montana Constitution’s definition of length-width compactness is unclear (and therefore unusable), there is a length-width measure of compactness called “LengthWidth” (LW) in the R package redist. This measure computes the ratio length/width of the north-south and east-west oriented bounding box containing the district. Length is assumed to be less than or equal to width so that the measure takes a score between 0 and 1. The smaller the score, the less compact the district by this measure. This is the compactness measure Iowa uses in their  \href{https://www.legis.iowa.gov/docs/code/42.4.pdf}{Redistricting Standard 4a}. Since there are two districts, each redistricting solution has two LW scores, one for each district. We will call the smaller of these two numbers the LW score of the redistricting solution. Figure \ref{fig:lw} contains the histogram of all LW measurements for redistricting solutions with population deviation $<0.1$. The red line is at 0.644 which is the LW measure of the adopted congressional districts.

\begin{figure}
    \centering
    \includegraphics[width = \linewidth]{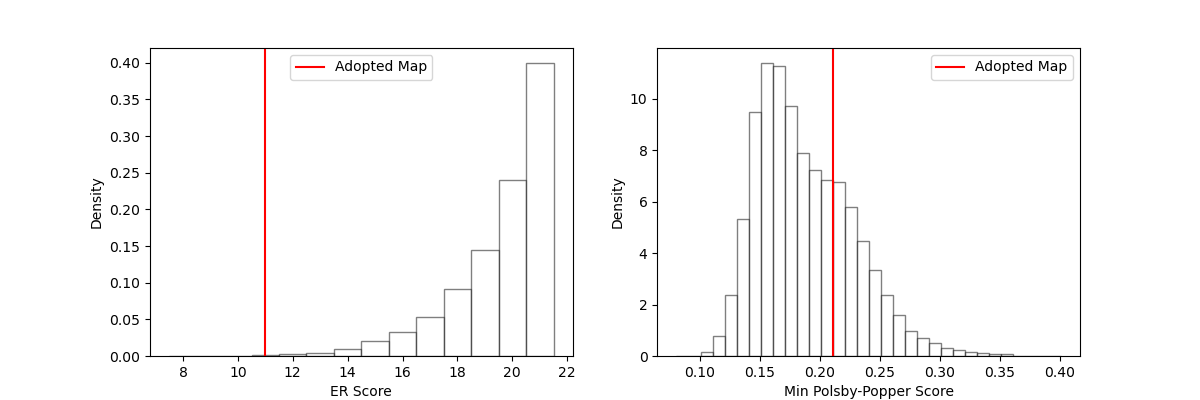}
    \caption{ER and Polsby-Popper scores of the 17,083 Enumerated Plans filtered for population deviation $<0.03$ and ER score $<22$.}
    \label{fig:ER+PBP}
    %Made with the CD GerryChain.ipynb file
\end{figure}

The final compactness measure that we mention here is the often used Polsby-Popper (PbP) score. Given a geographical region $R$, the Polsby-Popper score of $R$ is $\mathrm{ PbP}(R) = 4\pi A(R)/P(R)^2$ where $A(R)$ is the area of the region $R$ and $P(R)$ is the perimeter of $R$. Notice that $0\leq\mathrm{ PbP}(R)\leq1$ for all regions $R$ and under this measure a circle has $\mathrm{ PbP}$ score of 1. Regions with smaller scores are less compact by this measure.  The two districts in the adopted Montana congressional map have  PbP scores of 0.21 and 0.53. We consider both the minimum  PbP score across the two districts and the mean  PbP score. 
The min  PbP scores among all plans with population deviation $<0.03$ and ER score $<22$ are shown in Figure \ref{fig:ER+PBP}. As you can see, the adopted map scores better (smaller) on the ER distribution of plans than it does on the distribution of min Polsby-Popper scores, but by both considered measures is a very compact map. The relationship between min PbP score and ER score is illustrated by scatter plot in Figure \ref{fig:PbPvsER}. The figure shows that plans with a fixed ER score can have a wide range of PbP scores and vice versa.

\noindent
  \begin{minipage}[b]{0.48\textwidth}
    \centering
    \includegraphics[width = 1\linewidth]{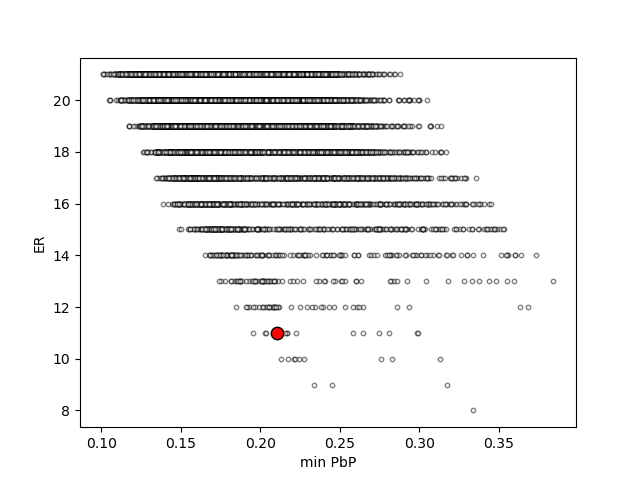}
    %\rule{6.4cm}{3.6cm}
    \captionof{figure}{Min PbP vs. ER score on the 17,083 enumerated plans. The adopted map is in red.}
    \label{fig:PbPvsER}
  \end{minipage}
  \hfill
  \begin{minipage}[b]{0.48\textwidth}
    \centering
 %   Scatter plot was made in 
 % CD GerryChain-reading in graph and df and analyze-neat.ipynb
\begin{tabular}{|c|p{.8in}|}
\hline
ER Restriction &  \# Redist 

Solutions\\
\hline\hline
\centering{None} &\hfill 30,011,928\\%95,259,204\\
\hline
\centering{$<22$} &\hfill 17,083\\
\hline
\centering{$<17$} & \hfill 1,206\\%936,080,801\\
\hline
\centering{$<12$} & \hfill 28\\%936,080,801\\
\hline
\end{tabular}
% \vspace{.35in}
\captionof{table}{Number of redistricting solutions from MT\_122 graph with population deviation $< 0.03$ and filtered by Edges Removed (ER) score. The adopted plan has an ER score 11.}
\label{table:ER}
% computed in CD GerryChain-reading in graph and df and analyze-neat.ipynb
    \end{minipage}

Since there are only 17,083 maps drawn along county lines with population deviation $<0.03$ and with ER score $<22$, it is easy to enumerate them and compare the adopted map's political outcomes to those in the enumerated ensemble of maps. This is done in the next section. 
 \section{Political leanings of Redistricting solutions}
\label{sec:politics}
 
Throughout this section we will compare the political outcomes of the adopted map with the political outcomes of all 17,083 possible ways to draw the congressional districts with reasonable compactness (ER less than 22) and population deviation (less than $.03$). Political outcomes are calculated assuming that the votes cast are identical to those of a previous state-wide race between 2016 and 2020 in both the location the vote occurred and whether the voter voted for the Democratic or Republican Party. We used the python package MAUP to proportionally allocate the votes to the correct district using 2020 census population counts. There are 10 state-wide races in our analysis. These are the same 10 state-wide races the DAC chose as applicable to political calculations in redrawing the state legislature 
\begin{wrapfigure}{r}{.4\linewidth}
    \includegraphics[width = \linewidth]{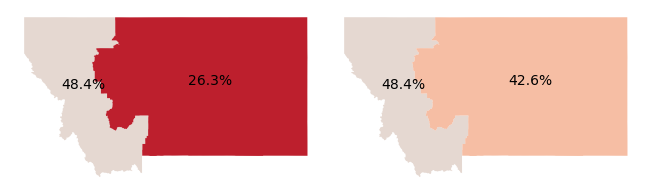}
    \caption{Percent of voters who voted Democratic (left) and Democratic or Independent (right) in the 2022 Congressional election.}
    \label{fig:pct_dem2022Cong}
\end{wrapfigure}
 in 2022-2023. We also include the 2022 Montana congressional election which occurred after the congressional map was adopted. Since the 2022 congressional election was not a statewide race (in fact the Independent candidate received 22\% of the vote in the Eastern District) the reader should be cognizant of this difference.

\begin{figure}[h]
 \includegraphics[width= \linewidth]{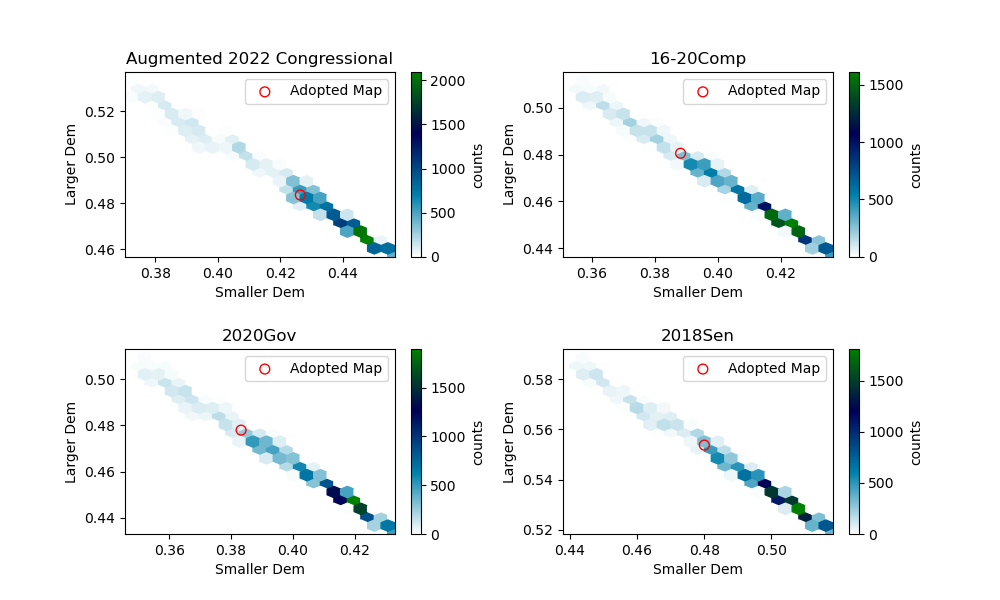}
 \caption{Heat map of $(x,y)$ where $x$ and $y$ are the proportion of voters who voted democratic in each of the two districts ordered so that $x\leq y$. These are calculated over all 17,083 redistricting solutions with ER $<22$ and population deviation $< 0.03$.}
 \label{fig:heatmap}
\end{figure}

Table \ref{fig:PctDemActual} shows the proportion of voters in the adopted eastern and western districts who voted for the Democratic Party candidate out of those who voted for either the Democratic or Republican Party candidate from each of these 10 elections. In addition, the figure includes the 2022 Congressional election and a composite election called `16-20Comp'. The composite election data was taken from \href{https://davesredistricting.org/maps#aboutdata}{Dave's Redistricting App} where the composite vote formulation can be found. From the elections prior to 2022 one can see that the eastern district was very likely to vote for the Republican Party candidate while the western district was more likely to be competitive. This held true for the 2022 Congressional elections in which the Eastern district  had 26.3\% of voters vote for the Democratic candidate and the Western district had 48.4\% vote for the Democratic candidate (again - considering only the republican and democratic candidates, thus ignoring all Libertarian and Independent Party votes). However, it should be noted that in the eastern district an additional $21.9\%$ of voters voted for the Independent Party candidate. If you combine the Independent Party voters with the Democratic Party voters (a big assumption), then the percentage of voters who voted Democratic or Independent is $42.6\%$ (see Figure \ref{fig:pct_dem2022Cong}). 
%The Libertarian vote was $1.4\%$ of the total vote in the eastern district and is ignored in this calculation.  
We will refer to the combined Democratic and Independent vote as the ``Augmented 2022 Congressional" election (see e.g., Figure \ref{fig:heatmap}). 

In Figure \ref{fig:heatmap} a scatter plot is used to compare the proportion of voters voting for the Democratic Party in the adopted congressional map with the 17,083 other possible maps with ER score $<22$ and population deviation $< 0.03$ discussed in section \ref{sec:compact}. In each of the four elections shown we assume that the voters who voted for the Democratic or Republican Party candidate vote the exact same way and vary the district lines among the 17,083 possibilities. The adopted map is seen to be an outlier in the scatter plot in that the proportion of voters voting democratic in the district with the smaller proportion (the horizontal axis) is less than most of the possible redistricting plans. Meanwhile, consider the proportion of voters voting democratic in the district with a larger proportion of voters voting democratic (the vertical axis). The adopted plan has a larger proportion than most of the other possible redistricting plans. 

\begin{figure}[h]
     \centering
     \includegraphics[width = \linewidth]{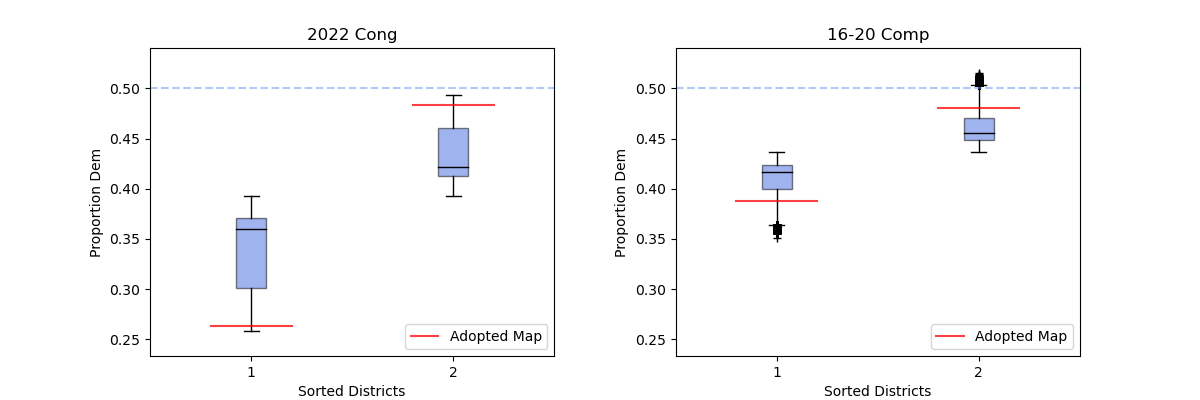}
     %Image made in CD GerryChain-reading in graph and df and analyze-neat.ipynb on 8/28/23
     \caption{Box plot of the smaller (Dist 1) and larger (Dist 2) proportion Democratic for each of the 17,083 redistricting solutions with ER $<22$ and population deviation $ <0.03$.}
     \label{fig:boxplot2}
 \end{figure}

Another way to look at this data is by box plot which can be found in Figure \ref{fig:boxplot2} for the 2022 Congressional election and the compilation election. The Adopted map is shown as the longer light red line and also presents as an outlier among the set of possible redistricting maps. Graphs for all 12 elections can be found in Figure \ref{fig:boxplot_dem} where the light red line occurs outside of the middle half of the distribution in each election.

\begin{figure}[h]
    \includegraphics[width = \linewidth]{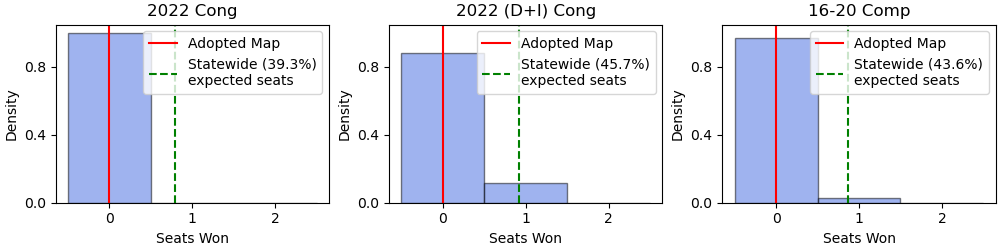}
    \caption{Of the 17,083 possible redistricting maps 1,970 of them elect one Democratic candidate if  you combine the Democratic and Independent Party votes in the eastern district.}
    \label{fig:dem_ele_small}
\end{figure}

\begin{figure}[h!]
     \centering
     \includegraphics[width = .83\linewidth]{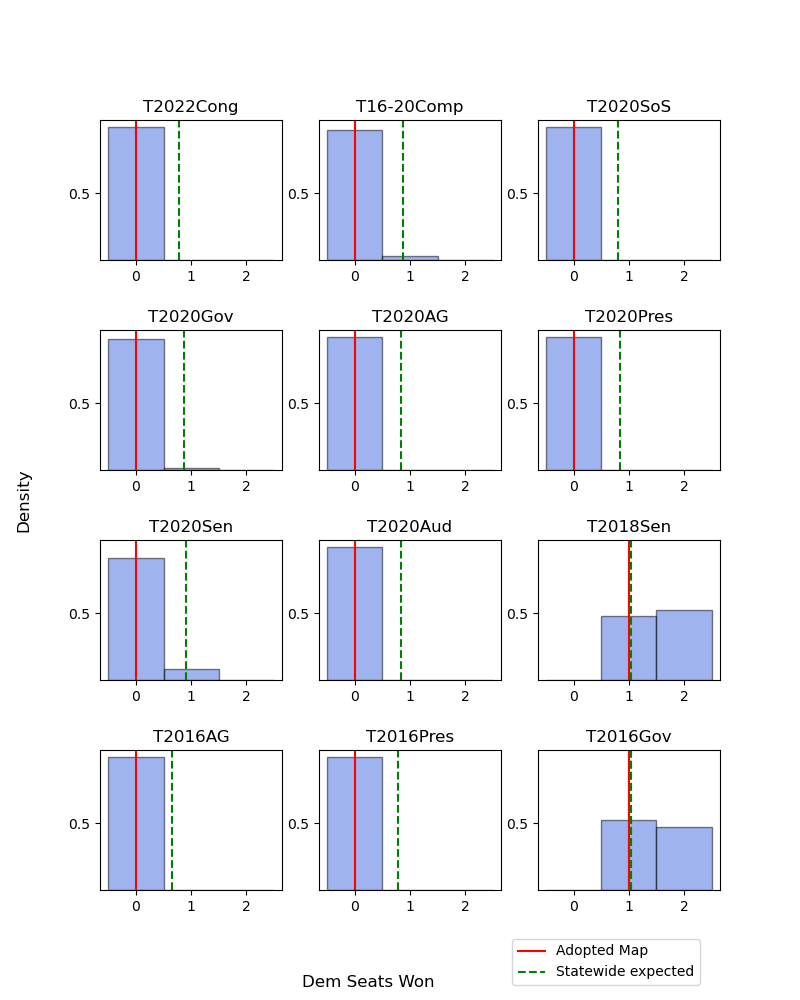}
     \caption{Density plot for the number of democrats elected in each of the 17,083 possible redistricting plans with ER $<22$ and population deviation $ <0.03$ for the 10 statewide races considered by the DAC plus the compilation election. The green dashed line is at twice the statewide percentage of voters who voted democratic, hence the number of democrats who would be elected for state-wide proportional representation. Though this is not an integer, it gives an idea of how far the adopted map and the ensemble of maps are from statewide proportionality}
     %created with CD GerryChain-reading in graph and df and analyze-neat
     \label{fig:dem_wins_enum}
\end{figure}

The last way we will look at how the adopted map compares to the set of all redistricting solutions politically is to look at the number of Democratic candidates elected. This can be found in Figure \ref{fig:dem_wins_enum} for all 12 elections and in Figure \ref{fig:dem_ele_small} for the 2022 election and the compilation election. In these figures the adopted map is indicated by the red vertical lines. In all but the 2018 Senator race,  of the 12 elections considered the red line correlates with the majority of the redistricting solutions, showing that the adopted map is not an outlier when considering the number of democrats elected.

Notably, using the voter data from the 2022 Congressional Election, none of the 17,083 possible redistricting maps elect a democrat. However, as can be seen in the middle figure of  Figure \ref{fig:dem_ele_small}, combining the Democratic and Independent Party votes shows that there are ways (in fact 1,970 of the 17,083 possible redistricting maps) to draw the maps so that one democrat would have been elected under the (Dem+Indep) assumption. This is assuming that each voter votes exactly as in the 2022 Congressional election for the fictitious Democratic and Republican candidates. In Figures \ref{fig:dem_ele_small} and \ref{fig:dem_wins_enum} the green dashed lines are at twice the statewide proportion of voters who voted democratic. This is the expected number of democrats who should be elected for statewide proportionality. Of course it is impossible to elect a non-integer number of democrats, so this can only be used as a measure of how far from proportionality the number of democrats elected is. In both the 2016 Governor and 2018 Senate elections, proportionality is nearly reached and around half of all possible maps elect two democratic candidates.

\section{Comparison of Sampling Methods}
\label{sec:sampling}
Montana's congressional districts provide a rare situation in which all possible redistricting maps that follow county lines can be enumerated. This is atypical as in other states or for other types of elections, there are too many districts for enumeration to be practical. To better understand these innumerable spaces, redistricting specific sampling and analysis algorithms have been developed. Two of these are the R package \href{https://cran.r-project.org/web/packages/redist/index.html}{{\it redist}}, which uses SMC sampling techniques developed in \cite{smc}, and the Python package \href{https://mggg.github.io/GerryChain/}{GerryChain} which uses MCMC algorithms developed in \cite{Recomb}. In this section we use these two packages to sample from the set of all possible redistricting plans. Then, we compare the sampled plans to the enumerated plans to see how the sampling techniques align with the true distribution of population deviation, compactness and political outcomes.

\begin{table}[ht]
    \centering
    \begin{tblr}{
  colspec = {|p{.73in}|p{.8in}|p{.7in}|p{2.98in}|}, rowspec = {Q[m]Q[m]Q[m]Q[m]Q[m]Q[m]},
}
        \hline
         {\bf Ensemble (Fig)} &{\bf Package}&{\bf Number (Unique)}&{\centering \bf Description}  \\
         \hline
         \centering 0&Redist (enumpart)&17,083 (17,083)&hard\_constraints = ER $<22$ \& pop\_dev$<.03$
         \\
         \hline
         %'mt_out_122_3pct_22ER_plans.dat'
         \centering 1(\ref{fig:ens_1})&Redist (SMC)&5,325 
         
         (776)&hard\_constraints = ER $<22$ \& pop\_dev$<.03$ \\
         \hline
         %CD_MCMC-3
         \centering 2(\ref{fig:ens_2})&GerryChain (ReCom)&100,000 (5,222)&hard\_constraints = ER $<22$ \& pop\_dev$<.03$ 
         
         edge\_accept = always accept

         One starting seed (adopted map)
         \\
         \hline
         %CD_MCMC-3-edgetry-4
         \centering 3(\ref{fig:ens_3})&GerryChain (ReCom)&100,000 (3,045)&hard\_constraints = ER $<64$ \& pop\_dev$<.4$ 
         
         edge\_accept = (ER $<22$ \& pop\_dev$<.03$ \& otherwise accept with prob .05)
         
         One starting seed (adopted map)\\
         \hline
         %CD_MCMC-RS-4
         \centering 4 (\ref{fig:ens_4})&GerryChain (ReCom)&100,000 (5,177)&hard\_constraints = ER $<22$ \& pop\_dev$<.03$ 
         
         edge\_accept = always accept

         5 starting seeds (adopted map plus 4 randomly chosen seeds from Ensemble 0)\\
         \hline
         
    \end{tblr}
    \caption{Ensembles 1 -- 4 were created from sampling software to compare them with the set of enumerated maps in Ensemble 0.}
    \label{table:ensemble_table}
\end{table} 

To generate our samples of possible plans, known as ``ensembles'', we use the same setup as in the enumerated plans. That is, we start with 55 intact counties and the Pondera split, giving 57 ``counties'' in total. Using {\it redist} we generated an ensemble of 6,000 plans (obtained from 2  independent runs) with a population deviation constraint of $ <0.03$ using the SMC function. The adopted congressional map has an ER score of 11, so we filtered this ensemble for ER scores less than 2 times the adopted plan; $<22$. These plans are called Ensemble 1 in Table \ref{table:ensemble_table}. The SMC run resulted in 5,325 plans which should be a sufficient sample as in \cite{mccartan2022simulated}. 

Using the Python package GerryChain and its MCMC algorithm for generating ensembles, we tested three sets of constraints and starting seeds. These are summarized in Table \ref{table:ensemble_table}. We experimented with a Metropolis-Hastings acceptance function in Ensemble 3 while Ensembles 2 and 4 always accepted the proposed plan. Each of the three GerryChain ensembles use the ReComb proposal (see \cite{Recomb}).

Ensembles 2, 3 and 4 each contain 100,000 plans. Ensemble 2, was generated with a population deviation constraint of $.03$ and ER score $<22$. Ensemble 3 had population deviation constraint  $ <0.4$ and ER scores $<64$. The accept function was `always\_accept' if both ER $<22$ and population deviation $ <0.03$ and otherwise accept with probability 0.05. Ensembles 1, 2, and 3 use the adopted map as the sole starting partition. Ensemble 4 was generated with a population deviation constraint of $.03$ and ER scores $<22$ where the accept function was `always\_accept'. The difference in Ensemble 4 is that there are 5 starting maps consisting of the adopted map and 4 randomly generated maps which were taken from Ensemble 0.

Since our ensembles contain repeat plans, we filter for the unique plans and list the number of unique plans in each ensemble in Table \ref{table:ensemble_table}. Both the full and unique ensembles are used for comparison to the enumerated plans and the adopted congressional map in the graphics that follow. In the next 2 subsections, our analysis will focus on Ensemble 2 since we feel the distribution aligns the best with the enumerated plans, but detailed graphics for each ensemble can be found in Figures \ref{fig:ens_1} - \ref{fig:ens_4}.

\subsection{Population Deviation \& Compactness}
\begin{figure}[h]
\centering
\includegraphics[width= \linewidth]{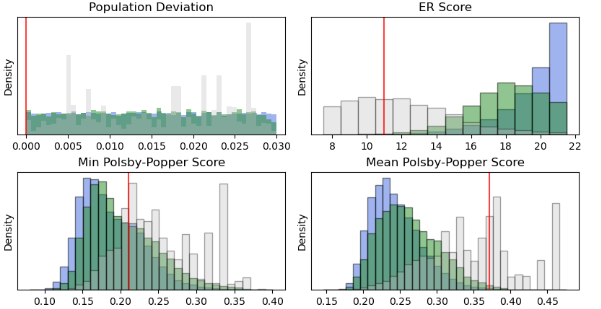}
%Images made in Erin Neat Figs.ipynb cut in half by snip tool
\caption{Population deviation, ER score, Polsby-Popper min, and Polsby-Popper mean from the 17,083 enumerated plans (blue), the full 100,000 plans (gray) and the 5,222 unique plans (green) from Ensemble 2. All plans have ER $<22$ and population deviation $ <0.03$. The figure is reproduced in Figure \ref{fig:ens_2} in the appendix using a kernel density estimate plot instead of a histogram.}
\label{fig:comp_results}
\end{figure}

Figure \ref{fig:comp_results} displays the population deviation and three compactness histograms comparing the full (gray) and unique (green) plans from Ensemble 2 and the enumerated plans (blue). In each sub-figure the adopted congressional map is indicated by a red vertical line. 
The three compactness measures shown are ER score, minimum Polsby-Popper (PbP), and mean PbP. 

The distribution of ER scores on the full Ensemble 2 is skewed further to the left (more compact) than the true distribution (blue). This is not a surprise as the ReComb proposal uses spanning trees to draw districts. In particular, the way the ReComb algorithm generates new plans is as follows; it combines two districts, randomly draws a spanning tree on the combined region (a spanning tree is a connected subgraph which includes all vertices and a minimal number of edges so that the subgraph has no circular paths) and then cuts an edge in the tree so that the two resulting districts have acceptable population deviation. If there is more than one edge that has acceptable population deviation it randomly chooses one such edge. Since Montana only has 2 districts, the algorithm draws a tree on the full MT\_122 map at each iteration of the chain. 

Multiple one-cut trees can give the same partition of counties into two districts, so the more one-cut trees that result in a given partition the higher the probability that that partition is proposed in the algorithm. Following \cite[\S5]{Recomb}, for any graph $G$ define $\sp(G)$ to be the number of spanning trees on $G$. This number is easily calculated using Kirchhoff’s matrix-tree theorem. Let $P=V_1\cup V_2$ be a partition of the graph nodes into two connected induced sub-graphs, i.e., a proposed redistricting plan. Define $\sp(P)$ to be the number of trees on the full graph for which you can cut an edge and get the partition $P$. Namely, 
\[\sp(P)=\sp(V_1)\cdot \sp(V_2)\cdot \mathrm{ER}(P)\] where $\mathrm{ER}(P)$ is the edges removed score. Let $E$ denote the enumerated set of plans in Ensemble 0 (pop\_dev $<$.03 and ER $<$22) and let $P \in E$. The probability that the MCMC ReComb algorithm with hard constraints pop\_dev $<$.03 and ER $<$22  will draw $P$ and the probability that a plan with ER score $C$ is drawn are respectively given by 
\[\mathrm{pr}(P) = \frac{\sp(P)}{\sum_{Q \in E} \sp(Q)}\hspace{.5in}\mathrm{and} \hspace{.5in} \mathrm{pr}(P\,|\,P \in E_C) = \frac{\sum_{P \in E_C}\sp(P)}{\sum_{Q \in E} \sp(Q)}\]
where $E_C$ is the subset of $E$ containing those plans with ER score $C$. These probabilities are easily computed and exhibit the relationship between ER score and the probability of being proposed in the MCMC ReComb algorithm.

\begin{wrapfigure}{l}{.5\linewidth}
    \includegraphics[width = \linewidth]{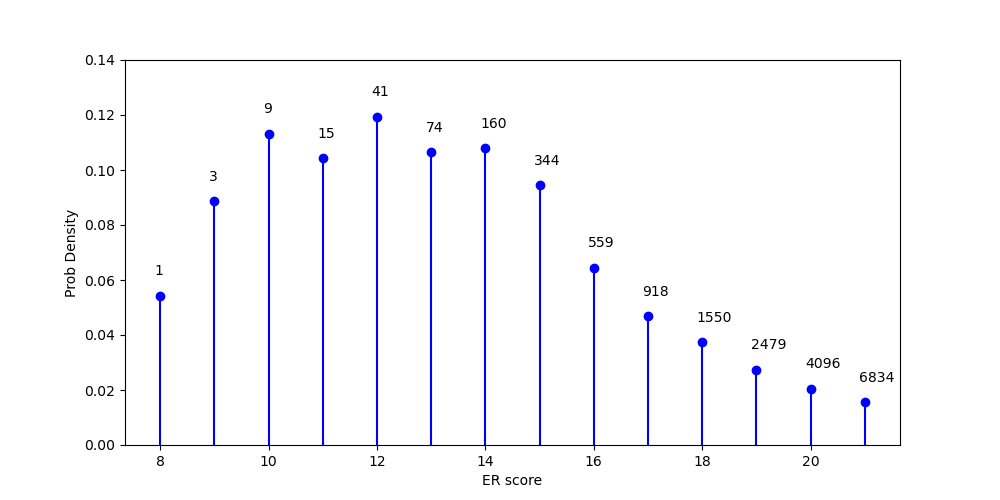}
    \caption{Probability distribution for tree drawing in MCMC ReComb proposal by ER score. The numbers above the bullets are the number of plans from the enumerated ensemble that have that ER score.}
    \label{fig:ERvProb}
\end{wrapfigure}

Figure \ref{fig:ERvProb} plots $\mathrm{pr}(P\,|\, P \in E_C)$ for each ER score $C$ occurring in $E$ with the number above each point the number of plans in the enumerated ensemble that has that ER score. So, for example, there is only one plan with ER score 8 and that one plan has a probability of around 0.054 of being drawn in the MCMC Recomb proposal. This explains the oversampling of smaller ER scores and explains the shape and location of the distribution of the full ensemble (gray) ER scores shown in Figure \ref{fig:comp_results}. The unique (green) ER scores in the figure are more representative of the true distribution. It is important to note here that the authors of \cite{Recomb} clearly state that their objective is not to sample uniformly from the distribution of plans because of the shear enormity of plans that are very non-compact and the impossibility of wading through these to sample those that are more compact. One possible concern is that comparing the adopted plan (red line) to the full (gray) distribution one would conclude that the plan is not all that compact and that there are many more compact plans. On the other hand, if one uses the enumerated (blue) or unique (green) distribution the adopted plan appears as an outlier in terms of how compact it is. In addition, including samples with higher ER scores may be important as there is the possibility that less compact regions allow for more state-wide proportionally representative plans or plans that at least change the political landscape significantly. Excluding the somewhat less compact plans may signify to the public or map drawers that there are no (or not many) ways to draw the map resulting in a different political outcome. 

Larger PbP scores indicate more compact districts, so again, the full Ensemble 2 plans (gray) skew toward better compactness scores than the true distribution of enumerated plans (blue). The unique plans (green) are again a better sample of the enumerated distribution of PbP scores.  The other ensembles from Table \ref{table:ensemble_table} had similar results. In particular, using 5 starting seeds in Ensemble 4 does not change the shape of the distributions from Ensemble 2 (compare Figures \ref{fig:ens_2} and \ref{fig:ens_4}) as was expected from the results in \cite{Recomb} showing very fast mixing in the ReComb algorithm and the large number of steps in our ensembles.

The population deviation sub-figure in Figure \ref{fig:comp_results} demonstrates that the unique (green) plans from Ensemble 2 align very well with the population deviations of the enumerated plans (blue). The full (gray) plans from Ensemble 2, however, have unnatural oversampling of certain population deviations. The discussion above on the MCMC ReComb proposal makes it likely that the over sampling of plans with the lowest ER scores causes the unusual population deviation sampling seen in the image.

The SMC algorithm for producing ensembles also relies on drawing spanning trees (\cite{smc}) so it is not a surprise that in Figure \ref{fig:ens_1}  compactness and population deviation behave similarly to that of the Recomb algorithm.

\subsection{Political Outcomes}
\begin{figure}[h]
\centering
\includegraphics[width= \linewidth]{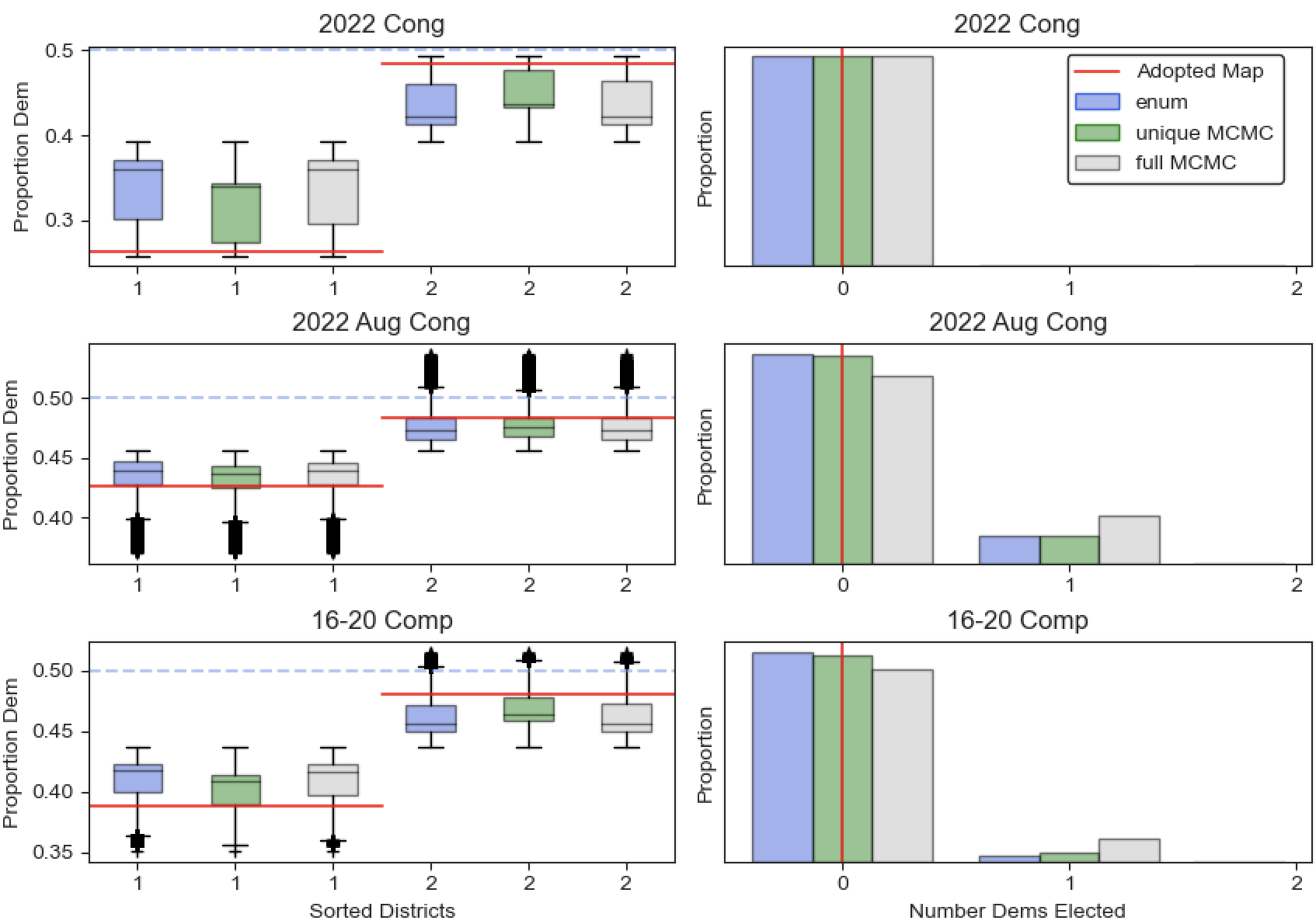}
%Images made in Erin Neat Figs.ipynb cut in half by snip tool
\caption{Comparison of the box plots of Democratic Party proportion and histograms of democratic party wins from 17,083 enumerated plans, full 100,000 MCMC plans and unique 5,222 MCMC plans of Ensemble 2. All plans have ER $<22$ and population deviation $ <0.03$.}
\label{fig:elec_win_results}
\end{figure}

\begin{figure}[h!]
     \centering
     \includegraphics[width = .92\linewidth]{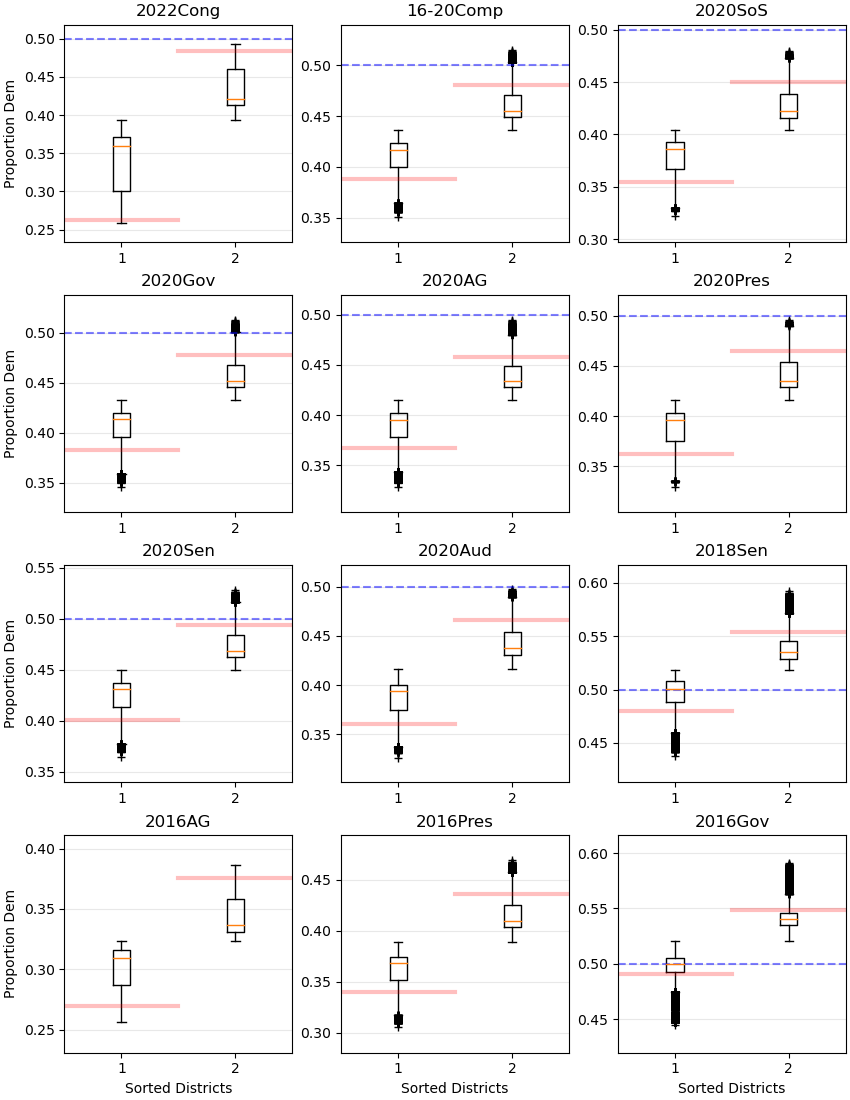}
     \caption{Box plot of the smaller (Dist 1) and larger (Dist 2) proportion Democratic for each of the 17,083 redistricting solutions with ER $<22$ and population deviation $ <0.03$.}
     %figure is from file CD GC-Erin Edit.ipynb
     \label{fig:boxplot_dem}
\end{figure}

In Figure \ref{fig:elec_win_results} we recreate in blue the box plots from Figure \ref{fig:boxplot_dem} which label District 1 as the district with smaller proportion of voters for the Democratic Party and District 2 as the larger proportion and run over all of the enumerated plans for the three given elections. Alongside these box plots are the corresponding box plots for the full (gray) and unique (green) plans in Ensemble 2. On the right hand side of the figure are the corresponding histograms of the number of democrats elected. The plots are shown for three elections: 2022 Congressional, 2022 Augmented Congressional, and 2016-2020 Compilation. The red horizontal line denotes the adopted map and the blue dashed horizontal line is at 0.5.

The full and unique plans in Ensemble 2 look similar to the enumerated plans with the full plans (gray) trending closer to the blue box plot this time. Comparing the blue boxes to the green boxes shows the effect of limiting plans to those with lower ER scores. Though it is slight, in the 16-20 Comp sub-figure of Figure \ref{fig:elec_win_results} it seems that sampling from more compact districts (green boxes) seems to widen the spread between the proportion Democratic in the two districts and the adopted plan is closer to the IQR. In terms of the number of democrats elected the unique plans are closer to the enumerated plans. 

In Figure \ref{fig:ens_3} the same graphs are given for Ensemble 3. In Ensemble 3 the MCMC algorithm was set to accept with probability 0.05 plans with ER$>$21 or population deviation $>$.03. You can see these non-conforming plans in the population deviation sub-figure. Accepting these plans changed the box plots, incorrectly suggesting, for example, that there are plans that elect democrats given the voting patterns of the 2022 congressional election. They also put the adopted congressional map in the IQR in District 1 for the 2022 Augmented Congressional election and the 16-20 Comp election when it is outside of IQR for the true (blue) distribution. %This shows that one needs to be careful in using the sampling software in drawing specific conclusions about the range of potential plans.

\section{Discussion}
\label{sec:discussion}
The enumeration of all possible redistricting plans along county lines can be used for the analysis of Montana's adopted congressional map and for the comparison of popular sampling methods. Montana's adopted congressional map is more compact than most of the enumerated maps in both its ER score and its PbP score (Figs. \ref{fig:comp} and \ref{fig:ER+PBP}). Although by each compactness score considered here the adopted map is an outlier, the DAC values compactness for practical and aesthetic reasons and is unlikely to contemplate compactness in comparison with other plans. For political outcomes the analysis on the adopted map is more mixed. On one hand the map seems to be an outlier when considering proportion Democratic for each district (Figs.\ref{fig:heatmap} and \ref{fig:boxplot2}). On the other hand it is not an outlier in terms of the number of Democrats elected (Fig. \ref{fig:dem_ele_small}).

\begin{figure}
    \centering
    \includegraphics[width = .95\linewidth]{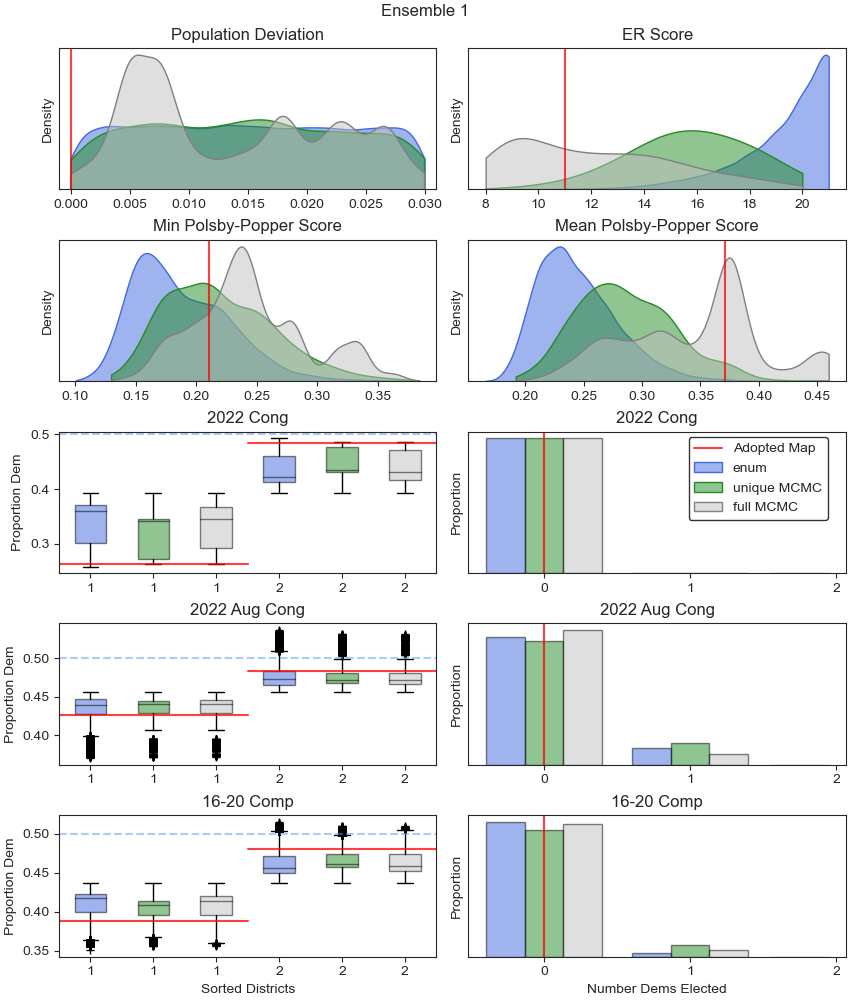}
    %image made in CD GerryChain-SMC figure.ipynb
    \caption{Ensemble 1 was created with the SMC function in the R package {\it redist} algorithm SMC. The full ensemble has 5,325 plans, 776 of which are unique. Hard constraints of ER $<22$ and pop\_dev $ <0.03$. The adopted map served as the one starting partition. The top four plots are kernel density estimate plots of the corresponding histograms for ease of viewing.}
    \label{fig:ens_1}
\end{figure}

\begin{figure}
    \centering
    \includegraphics[width = .95\linewidth]{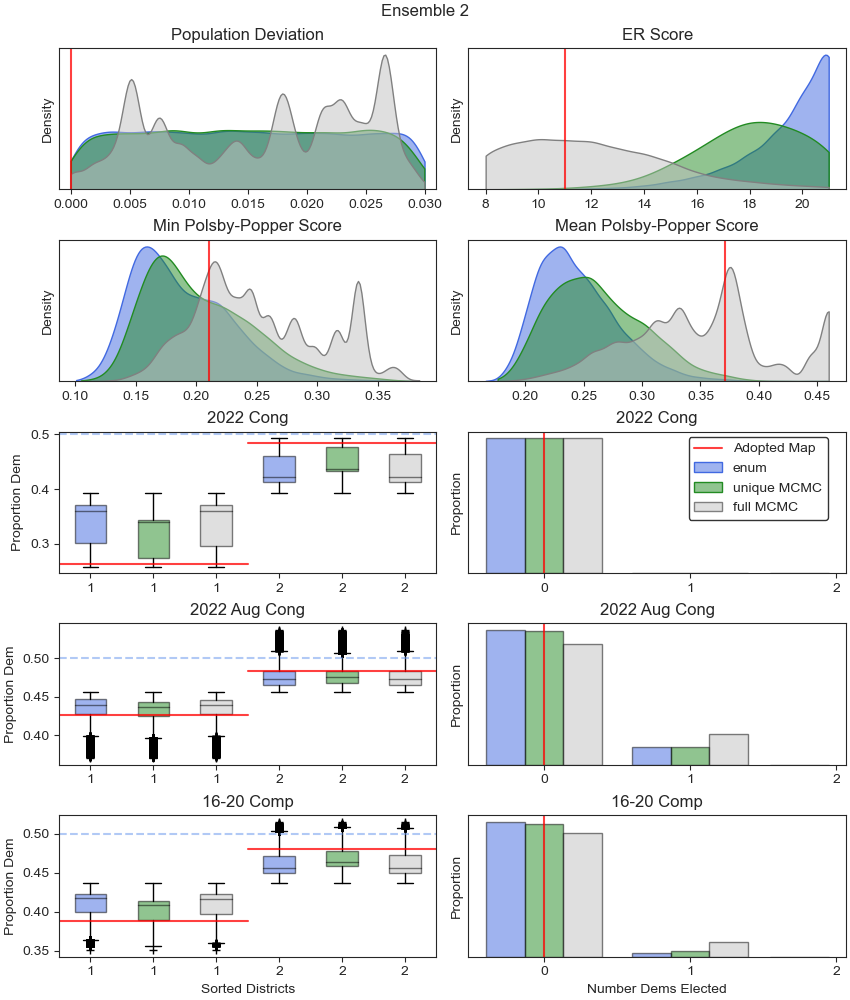}
    %image made in Erin Neat Figs.ipynb
    \caption{Ensemble 2 was created with the GerryChain proposal ReCom. The full ensemble has 100,000 plans, 5,222 of which are unique. The hard constraints in this MCMC run were ER $<22$ and pop\_dev $ <0.03$. The accept function was `always\_accept'. The adopted map served as the one starting partition. The top four plots are kernel density estimate plots of the corresponding histograms for ease of viewing.}
    \label{fig:ens_2}
\end{figure}

\begin{figure}
    \centering
    \includegraphics[width = .95\linewidth]{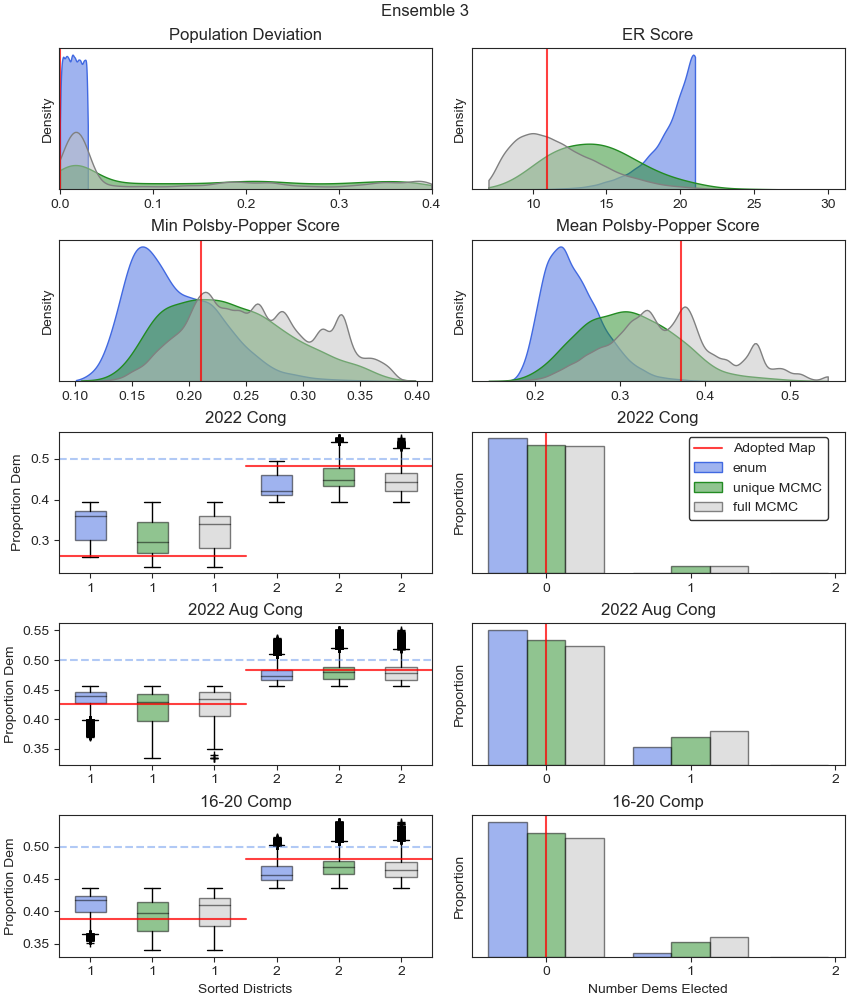}
    %image made in Erin Neat Figs.ipynb
    \caption{Ensemble 3 was created with the GerryChain proposal ReCom. The full ensemble has 100,000 plans, 3,045 of which are unique. The hard constraints in this MCMC run were ER $<64$ and pop\_dev $ <0.4$. The accept function was `always\_accept' if both ER$<22$ and pop\_dev$ <0.03$ and otherwise accept with probability 0.05. The adopted map served as the one starting partition. The top four plots are kernel density estimate plots of the corresponding histograms.}
    \label{fig:ens_3}
\end{figure}

\begin{figure}
    \centering
    \includegraphics[width = .95\linewidth]{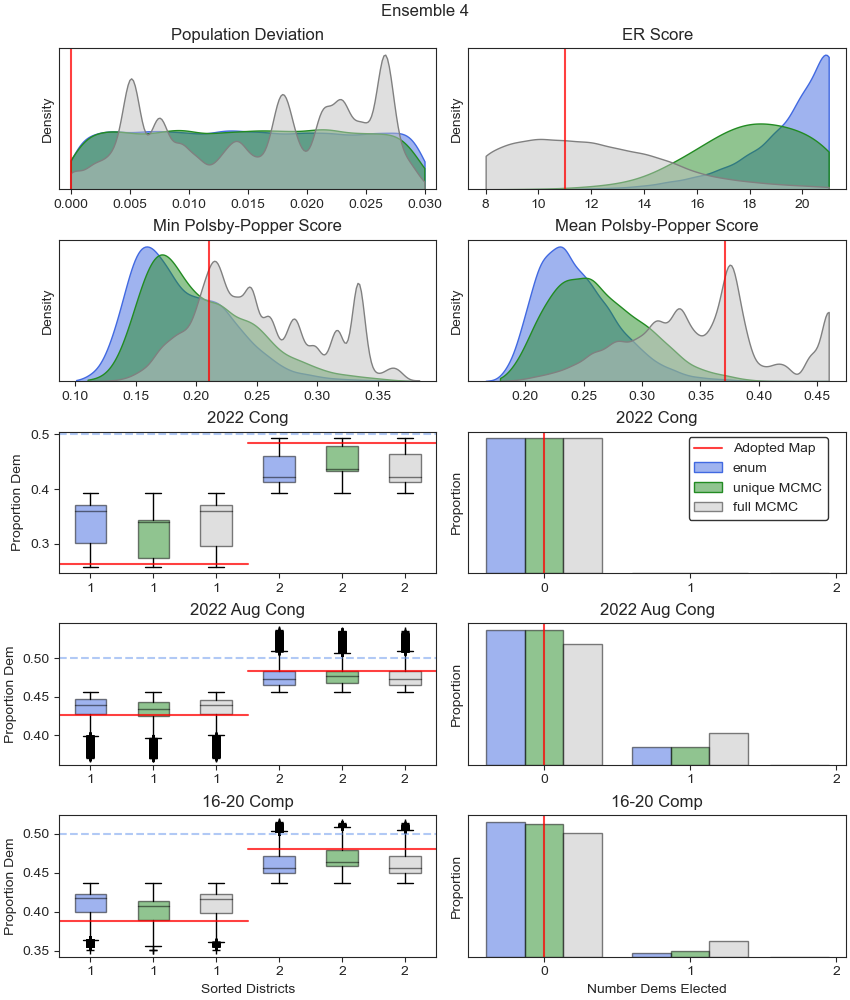}
    %image made in Erin Neat Figs.ipynb
    \caption{Ensemble 4 was created with the GerryChain proposal ReCom. The full ensemble has 100,000 plans, 5,177 of which are unique. The hard constraints in this MCMC run were ER $<22$ and pop\_dev $ <0.03$. The accept function was `always\_accept'. There were 5 starting maps consisting of the adopted map plus 4 randomly generated maps (taken from Ensemble 0). The top four plots are kernel density estimate plots of the corresponding histograms for ease of viewing.}
    \label{fig:ens_4}
\end{figure}

The SMC and MCMC sampling techniques that we tested gave compactness results that were not consistent with the enumerated plans (Figs.\ref{fig:ens_1},\ref{fig:ens_2},\ref{fig:ens_3},\ref{fig:ens_4}) as was to be expected from \cite{Recomb}. The full SMC and MCMC ensembles had higher densities of restricting plans that were more compact. Having reasonably compact districts is a goal of the commission so this particular result would not change whether or not the adopted map was acceptable, however it would change the perception of how the map fits in with the space of all possible maps. In particular, the adopted map would not appear unusually compact if compared only to these ensembles. The unique plans' (green) compactness scores aligned better with enumerated plans' compactness scores.

The political outcomes of the SMC and MCMC ensembles appear to align more closely with the enumerated maps. For Ensembles 2 and 4 (MCMC), the Democratic party proportions for the 16-20 Compilation election have the adopted map inside the IQR but outside the IQR for the enumerated maps. This can been seen in a more extreme way for Ensemble 3, which was generated using a Metropolis-Hastings accept function (Fig.\ref{fig:ens_3}).

Ultimately, the sampling techniques give slightly different results than the enumerated plans, however, as stated previously, enumerating all possible plans is unrealistic in most redistricting situations. When using the sampling tools, it is important to understand the distribution of compactness scores from which the samples are taken and the potential differences from the true distribution. 

\section{Acknowledgements}
This paper and these results stem from an undergraduate research project at the University of Montana. We thank Professor David Patterson and Professor Daryl Ford for their help with this project over the last two years. Thanks also to the Office of Research and Policy Analysis of Montana's Legislative Services Division for help acquiring, parsing and ultimately fixing some data that was posted on the DAC's website.  

Thanks also to the University of Montana's Office of Research and Creative Scholarship's  University Grant Program which helped fund the writing of this paper in Summer 2023. We also wish to thank the University of Montana's Department of Mathematical Sciences for support of students with Undergraduate Research Awards. The following students were Undergraduate Research Scholars and participated in the ``Mathematics of Redistricting Montana'' project over the last two years: Vivian Cummins, Ian Oberbillig, Noah Ryan \& Erin Szalda-Petree.

\section{Appendix of Tables and Figures}
\label{sec:figs_and_tables}

\begin{table}[h!]
    \centering
\begin{tabular}{b{.75in}b{.6in}b{.6in}||b{.7in}b{.6in}b{.6in}} \toprule \vfill {\bf Election} & {\bf Western District} & {\bf Eastern District}&\vfill {\bf Election} & {\bf Western District} & {\bf Eastern District} \\ \midrule 
2022Cong & \hfill0.484 & \hfill 0.263 &
2020Sen & \hfill 0.494 & \hfill 0.401\\
&&&&&\\ 
16-20Comp & \hfill 0.481 & \hfill 0.388 &
2020Aud & \hfill 0.466 & \hfill 0.361\\
&&&&&\\ 
2020SoS & \hfill 0.450 & \hfill 0.354 &
2018Sen & \hfill 0.554 & \hfill 0.480\\
&&&&&\\ 
2020Gov & \hfill 0.478 & \hfill 0.383 &
2016AG & \hfill 0.375 & \hfill 0.269\\
&&&&&\\ 
2020AG & \hfill 0.458 & \hfill 0.367 &
2016Pres & \hfill 0.436 & \hfill 0.340\\
&&&&&\\ 
2020Pres & \hfill 0.465 & \hfill 0.362 &
2016Gov & \hfill 0.549 & \hfill 0.490\\
\bottomrule\\ 
\end{tabular}
\caption{For the Western and Eastern adopted congressional districts the percent of voters who voted for the Democratic Party candidate out of those who voted for either the Democratic or Republican Party candidate only in each of the 10 elections used by the DAC plus the 2022 Congressional election and the 16-20 Compilation. Note that in 2022, 22\% of voters voted for the Independent Party candidate in the Congressional election in the Eastern District. %Including all independent voters with the democratic voters changes the percentage from 26.3\% to 42.6\%. 
These results can be found on the \href{https://electionresults.mt.gov/resultsSW.aspx?type=FED&map=CTY}{MT SOS page}.}
\label{fig:PctDemActual}
\end{table}
\newpage
\begin{table}
\begin{tblr}{colspec = {lllp{.8in}rr}, rowspec = {Q[b]Q[m]Q[m]Q[m]Q[m]Q[m]},}
\hline
 & \centering County 1 & County 2 & Shared Perim (km) & \% County 1 & \% County 2 \\
\hline
0 & Deer Lodge & Ravalli & 0.2 & 0.06 & 0.03 \\
1 & Custer & Garfield & 2.3 & 0.34 & 0.24 \\
2 & Glacier & EPondera & 5.4 & 0.78 & 1.12 \\
3 & Meagher & Sweet Grass & 5.8 & 0.90 & 1.10 \\
4 & Rosebud & Yellowstone & 7.1 & 0.70 & 1.05 \\
5 & Liberty & EPondera & 14.6 & 3.32 & 3.03 \\
6 & Golden Valley & Yellowstone & 16.1 & 3.81 & 2.40 \\
7 & Flathead & Missoula & 18.7 & 1.59 & 2.31 \\
8 & Deer Lodge & Jefferson & 18.8 & 4.85 & 3.72 \\
9 & Gallatin & Jefferson & 22.5 & 2.91 & 4.47 \\
10 & Chouteau & EPondera & 24.2 & 3.00 & 5.01 \\
11 & Fallon & Prairie & 25.8 & 5.72 & 4.54 \\
12 & Carbon & Park & 26.0 & 4.28 & 3.93 \\
13 & Prairie & Wibaux & 29.0 & 5.12 & 7.22 \\
14 & Park & Stillwater & 30.4 & 4.59 & 5.23 \\
15 & Gallatin & Meagher & 31.0 & 4.00 & 4.77 \\
16 & Big Horn & Powder River & 31.1 & 3.31 & 5.76 \\
17 & Flathead & Powell & 36.6 & 3.11 & 5.14 \\
18 & Petroleum & Rosebud & 37.4 & 5.84 & 3.74 \\\hline[dashed]
19 & Fergus & Wheatland & 39.8 & 4.34 & 10.78 \\
20 & Fergus & Musselshell & 42.1 & 4.59 & 8.41 \\
21 & Chouteau & Judith Basin & 47.4 & 5.88 & 9.19 \\
22 & Chouteau & Teton & 48.0 & 5.95 & 7.70 \\
23 & Richland & Wibaux & 49.1 & 8.44 & 12.19 \\
\hline
\end{tblr}
\caption{MT counties with shared border. Ordered in increasing length of shared perimeter. Edges corresponding to those counties above the dashed line are removed from the MT\_141 graph to form the graph MT\_122.}
\label{table:borderlengths}
\end{table}

\newpage

\newpage

\newpage
\newpage
\newpage
\newpage
\newpage
\clearpage

\bibliographystyle{plain} % We choose the "plain" reference style
\bibliography{refs.bib}
\end{document}